\documentclass[aps,pra,onecolumn,superscriptaddress,notitlepage,nofootinbib,longbibliography, colorlinks=true]{revtex4}
\usepackage{amsmath,amssymb,amsfonts,graphicx,float,times,psfrag}
\usepackage[pdftex]{color}
\usepackage{amsmath,bm}
\usepackage[colorlinks, linkcolor=blue, citecolor=blue,  breaklinks=true]{hyperref}
\usepackage{mathtools,amsfonts,mathptmx}
\usepackage[utf8]{inputenc}
\usepackage[T1]{fontenc}
\usepackage{xcolor}
\usepackage{braket}
\usepackage[titletoc,title]{appendix}
\usepackage[nameinlink,capitalize]{cleveref}

\usepackage[shortlabels]{enumitem}

\begin{document}

\title{Phonon-hopping-assisted entanglement and quantum complete synchronization in a magnomechanical system }

\author{E. Kongkui Berinyuy}
\email{emale.kongkui@facsciences-uy1.cm}
\affiliation{Department of Physics, Faculty of Science, University of Yaounde I, P.O.Box 812, Yaounde, Cameroon}

\author{P. Djorwé}
\email{djorwepp@gmail.com}
\affiliation{Department of Physics, Faculty of Science, 
University of Ngaoundere, P.O. Box 454, Ngaoundere, Cameroon}
\affiliation{Stellenbosch Institute for Advanced Study (STIAS), Wallenberg Research Centre at Stellenbosch University, Stellenbosch 7600, South Africa}

 \author{A. N. Al-Ahmadi}
 \affiliation{Department of Physics, College of Sciences, c University, Makkah 24382, Saudi Arabia}

 \author{H. Ardah}
 \affiliation{Department of Computer Sciences, College of Computer and Information Sciences, Princess Nourah bint Abdulrahman University, P.O.Box 84428, Riyadh 11671, Saudi Arabia
 	} 

\author{A.-H. Abdel-Aty}
\affiliation{Department of Physics, College of Sciences, University of Bisha, Bisha 61922, Saudi Arabia}

\begin{abstract}
Quantum entanglement and synchronization in cavity magnomechanical system has a wide range of applications in  modern quantum technologies. In this work, we propose a theoretical scheme to generate and enhance quantum entanglement and quantum complete synchronization through the interplay of cross-Kerr nonlinearity and mechanical parametric amplification (MPA) in a magnomechanical system. Our system is made of a magnonic mode that is simultaneously driving the acoustic phononic and the center-of-mass motion (CMM) phonon in a yttrium iron garnet sphere. The acoustic mode and the center-of-mass mechanical (CMM) mode are weakly coupled via the phonon hopping rate $J_m$. Moreover, the magnonic and phononic modes interact through cross-Kerr interaction, while the phononic mode is additionally driven via a mechanical Parametric Amplification (MPA). Without the  mechanical coupling ($J_m = 0$) and the MPA, the generation of entanglement among the subsystems requires a relatively strong effective cross-Kerr coupling. However, when phonon hopping and MPA are accounted, quantum entanglement can be generated even for modest values of the effective cross-Kerr coupling strength, revealing the key role of these interactions  in the engineering of quantum correlations in our proposal. Furthermore, we demonstrate that the coupled mechanical modes exhibit quantum complete synchronization arising from the interplay between direct phonon hopping and the magnon-mediated interaction, while the cross-Kerr nonlinearity provides an efficient means of controlling the synchronization behaviour. Our work suggests how robust and stable quantum correlations and quantum synchronization can be engineered in magnomechanical structures based on nonlinear interactions. These results are useful for modern quantum applications including quantum information processing, quantum communication, and quantum computational tasks.  
\end{abstract}

\pacs{ 42.50.Wk, 42.50.Lc, 05.45.Xt, 05.45.Gg}
\keywords{Entanglement, magnomechanics, cross-Kerr}  
\maketitle
\date{\today}

\section{Introduction}\label{intro}
Owing to the recent progress in quantum science, quantum correlations and quantum synchronization have become interesting resources for countless quantum applications. Considerable effort is being invested in engineering quantum correlations in wide variety of physical systems such as   Cavity Quantum Electrodynamics (CQED) ~\cite{Amgain2025}, superconducting circuits ~\cite{Meesala2024,Gove2025}, optomechanical structures ~\cite{Berinyu2025,Imara,Liu2025,Chen2025,Cheng2024,BERINYUY2026418939,Beriny2026}, and optomagnomechanical systems ~\cite{Fan2023,Hu2025}. These quantum correlations are crucial for number of quantum technologies including quantum communication~\cite{Li2017}, quantum information processing~\cite{Wendin2017}, metrology and sensing ~\cite{Djorwe2019,Xia2023,Huang2024,Djorw2024}, and quantum computing~\cite{Zidan2021}, among others. On the other hand, quantum synchronization provides a powerful framework for understanding coherent dynamics and collective behaviour in coupled quantum systems, with promising applications in quantum state control, and distributed quantum networks~\cite{Roulet2018,Rodrigues,Shenn2023}. Consequently, the simultaneous engineering of quantum entanglement and synchronization has attracted considerable attention in a wide range of hybrid quantum platforms~\cite{7hhn-zypy,Berinyuyy2026,PENG2025116473,MANJU2023129039}.

Among these platforms, hybrid systems involving magnonic modes have attracted increasing interests in the recent years, due to their rich spectrum of modes and remarkable degree of tunability~\cite{Walker}. These quasiparticles, which represent collective spin excitations in magnetic materials, offer a versatile platform for controlling and manipulating information at the quantum level. Their multimode nature enables complex interactions and coupling mechanisms, while their tunability achieved through external magnetic fields, geometry, or material engineering makes them particularly promising for applications in hybrid quantum systems, and quantum information processing~\cite{awschalom2021,Yuan,Liu}.  Moreover, magnonic structures have been extensively investigated through spherical yttrium iron garnet (YIG) magnets~\cite{Xu2023,Rao}, whose high spin density and ultralow damping make them an exceptional platform for coherent interactions. These properties have enabled the observation of strong coupling between magnonic modes and wide range of quantum entities including photons, phonons, and superconducting qubits ~\cite{Hou,Pan}. 
Owing to the strong coupling between modes, together  with the nonlinear interactions between them, these hybrid structures involving magnons are interesting for the generation of non-classical states and robust quantum correlations as recently reported in  Ref.~\cite{Huo2025}.  Nonlinearity plays a central role in these developments. In optomechanical systems with Kerr nonlinearity, it has been shown to influence the quantum synchronization of two mechanical oscillators~\cite{Qiao2018}. In magnonic systems, the Kerr effect has been exploited to produce strongly non-classical single-magnon statistics, as in magnon blockade in a spinning microwave resonator~\cite{Bhatt2025a}. Entanglement in magnomechanical settings has likewise been investigated, for example between two magnon modes mediated by the magnetostrictive interaction~\cite{Li2019}, and in opto-magnomechanical systems, where population and entanglement dynamics have been studied in the presence of a two-level emitter~\cite{Bhatt2025}.

These works address complementary aspects of the problem. Refs.~\cite{Qiao2018} and~\cite{Bhatt2025a} focus on Kerr nonlinearity, for synchronization of mechanical oscillators and for single-magnon blockade respectively, rather than for continuous-variable entanglement. Refs.~\cite{Li2019} and~\cite{Bhatt2025} concentrate on entanglement in magnomechanical and opto-magnomechanical systems, driven by the magnetostrictive interaction or by coupling to an emitter, and do not combine it with synchronization. Previously, robust dynamical and steady-state entanglement of phonons was engineered in a magnetic microsphere attached to a cantilever and subjected to a uniform bias field ($B_o$) and a weak gradient driving field ($H_d$)~\cite{Huo2025}. That scheme, like the Kerr-based and magnetostrictive studies above, treats only entanglement.

 Despite this progress, three open issues remain. Firstly, magnonic and magnomechanical entanglement schemes usually depend on a single nonlinear mechanism, and they often require a strong nonlinear coupling that is hard to reach in practice. In particular, the cross-Kerr interaction between magnons and phonons is intrinsically weak~\cite{Shen}, which limits the entanglement it can generate on its own. Secondly, the entanglement and synchronization are typically studied in separate systems, so it is unclear whether the same resources can create and control both. Thirdly, the role of mechanical coupling between distinct phonon modes, such as phonon hopping, combined with parametric amplification, has not been explored as a way to compensate for modest nonlinearity. This motivates the central question of our work: can additional, experimentally accessible mechanical interactions enable entanglement and synchronization in a magnomechanical system even when the effective cross-Kerr coupling is modest? A YIG sphere is a natural setting, since it hosts a magnon mode together with an acoustic phonon mode and the center-of-mass motion (CMM), whose mutual coupling can be engineered~\cite{Huo2025}. Motivated by these open issues, we investigate how phonon hopping and mechanical parametric amplification (MPA) can enhance quantum correlations and establish quantum complete synchronization in a magnon-phonon system with cross-Kerr nonlinearity.
 
Our benchmark system consists of a magnonic mode that is coupled to both acoustic phononic mode and the center-of-mass motion (CMM) phonon in a yttrium iron garnet sphere. Both acoustic and the phononic modes are weakly mechanically coupled, and this coupling is captured through a phonon hopping rate $J_m$. Moreover, a cross-Kerr nonlinearity mediates interaction between the magnonic and phononic modes, while the phononic mode is additionally driven by a Mechanical Parametric Amplification (MPA).  We found that, without both mechanical coupling and parametric amplification, the generation of  entanglement requires a relatively strong cross-Kerr interaction. However, when phonon hopping and MPA are taken into account, quantum entanglement can be achieved even for a modest strength of the effective cross-Kerr nonlinearity. This reveals how phonon hopping plays a key role in redistributing quantum correlations across the system, enabling efficient generation of both bipartite and tripartite entanglement. 
  Our findings reveal that the same nonlinear interactions responsible for enhancing quantum correlations can also be exploited to control coherent mechanical synchronization, thereby providing a versatile platform for simultaneously engineering entanglement and synchronization in hybrid magnomechanical systems. These results may have potential applications in quantum information processing, quantum sensing,  and quantum computational tasks.

The rest of our work is organized as follow. \Cref{sec:model} introduces the theoretical model and outlines the derivation of the dynamical equations. \Cref{sec:Re} delves into  numerical results and discussion. In \Cref{sec:purity}, we discuss quantum complete synchronization, while \Cref{sec:concl} concludes our work.

\section{Model and dynamical equations} \label{sec:model}
\begin{figure}[htp!]
	\centering
	\includegraphics[width=0.6\linewidth]{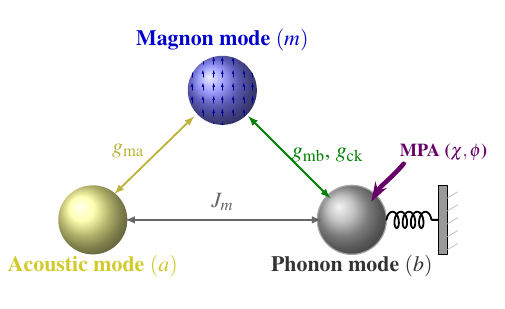}
	\caption{ Sketch of our benchmark system. A magnonic mode ($m$) that simultaneously drives  both phononic ($b$) and acoustic ($a$) modes, which are mechanically coupled through a phonon hopping rate $J_m$.  A cross-Kerr interaction couples the magnonic and the phononic modes, and a mechanical parametric amplification is driving the microsphere.}
	\label{fig:setup}
\end{figure}

Our benchmark system consists of a yttrium iron garnet (YIG) microsphere with a radius of $100 nm$ (phononic mode) that is mechanically coupled to the free moving-end of a cantilever beam (acoustic mode).  The mechanical coupling is captured through the phonon hopping $J_m$, while the microsphere is subjected to an external magnetic field consisting of a uniform bias component along the z-axis and a weak gradient driving field~\cite{Huo2025}. Moreover, we assume a cross-Kerr interaction between the magnonic and the phononic modes, while mechanical parametric amplification is driving the microsphere. Our proposal is sketched in \Cref{fig:setup}, and it is described by the following Hamiltonian ($\hbar=1$),
\begin{equation}\label{eq:Ham}
\begin{aligned}
\mathcal{H} &= \omega_m m^\dagger m + \omega_a a^\dagger a + {\omega}^\prime_b b^\dagger b + g_{ma} (m a^\dagger + m^\dagger a) + g_{mb} (m b + m^\dagger b^\dagger)+ J_m (a^\dagger b + a b^\dagger)
+ g_{ck} m^\dagger m b^\dagger b + \frac{\chi}{2}(b^{\dagger 2}e^{-i\phi} + e^{i\phi}b^2)\\& + i  (Em^\dagger e^{-i\omega_\ell} -E^* me^{i\omega_\ell}),
\end{aligned}
\end{equation}
where $m (m^\dagger)$, $a (a^\dagger)$, and $b (b^\dagger)$ are the annihilation (creation) operators for the magnonic, acoustic, and phononic modes, obeying the commutation relations $[\mathcal{O}_j, \mathcal{O}_j^\dagger] = 1$, $\mathcal{O}=a,b,m$. The resonance frequencies of the magnonic, acoustic and CMM modes are respectively, $\omega_{m}$, $\omega_{a}$ and  $\omega_b^\prime$. The first three terms of our Hamiltonian capture the free energies of the magnonic, acoustic, and phononic modes. The fourth, fifth and the sixth terms represent the coupling between the magnonic and the acoustic modes through the coupling strength $g_{ma}$, the coupling between the magnonic and the phononic modes through the coupling strength $g_{mb}$, and the coupling between the acoustic and the phononic modes through the coupling strength $J_{m}$, respectively. The seventh and the eighth terms represent the cross Kerr nonlinearity and the mechanical parametric amplification respectively. The last  terms  captures the driving field with amplitude $E$. The parameter $\chi$ is the parametric driving amplitude which comes from the modulation of the spring constant of the center-of-mass of the microsphere (CMM) at the frequency $2\omega_d$ with a phase $\phi$.

In the frame rotating at the frequency $\omega_{\ell}+\omega_d$, where $\omega_{\ell}$ denotes the electromagnetic driving frequency, we apply the rotating wave approximation (RWA) to the above Hamiltonian that yields, 
\begin{equation}\label{eq:hamil}
\begin{aligned}
\mathcal{H}&= \Delta_m\, m^\dagger m + \Delta_a\, a^\dagger a + \tilde{\omega}_b\, b^\dagger b  + g_{ma} (m a^\dagger + m^\dagger a) + g_{mb} (m b + m^\dagger b^\dagger) + J_m (a^\dagger b + a b^\dagger)+ g_{ck} m^\dagger m b^\dagger b + \frac{\chi}{2}(b^{\dagger 2} +b^2)+ iE(m^\dagger - m),
\end{aligned}
\end{equation}
where $\Delta_{m}=\omega_{m}-\omega_{\ell}$, $\Delta_{a}=\omega_{a}-\omega_{\ell}$, and $\tilde{\omega}_b=\omega_b^\prime-\omega_d$, are  the frequency detunings for the magnonic, acoustic, and the phononic modes. We would like to mention that the modulation phase has been neglected in our investigation ($\phi=0$). We will made the assumption, $\tilde{\omega}_b=\omega_b$ for a seek of simplicity. To diagonalize the quadratic term within our Hamiltonian, we introduce the squeezing transformation $S(r) = \exp\left[\frac{r}{2}\left(b^2 - b^{\dagger 2}\right)\right]$,
where the squeezing parameter is defined as
$r= \frac{1}{4} \ln \left(\frac{\tilde{\omega}_b + \chi}{\tilde{\omega}_b - \chi}\right)$. By using the Bogoliubov transformation $b_{s} = b \cosh r + b^\dagger \sinh r$, \cref{eq:hamil} becomes,
\begin{equation}
\begin{aligned}
\tilde{\mathcal{H}} &= \Delta_m^\prime m^\dagger m + \Delta_a a^\dagger a + \Delta_b b_s^\dagger b_s + g_{ma} (m a^\dagger + m^\dagger a) + \tilde{g}_{mb} (m^\dagger b_s + m b_s^\dagger) +  \tilde{J}_m (a^\dagger b_s + a b_s^\dagger)  + \tilde{g}_{ck}, m^\dagger m b_s^\dagger b_s + i E (m^\dagger - m),
\end{aligned}
\end{equation}
where we have defined the following modulated coefficients $\tilde{J}_m=J_m \cosh r$, $\tilde{g}_{ck}=g_{ck} \cosh 2r$, $\Delta_b = \sqrt{\tilde{\omega}_b^2 - \chi^2}$,  $
\tilde{g}_{mb}=-g_{mb}\sinh r$ 
 and $\Delta_m^\prime=\Delta_m+g_{ck}\sinh^2r.
$ 

\subsection{\label{sec:IIB}Quantum Langevin Equations} 

By using the Heisenberg equation, we can derive the following set of Quantum Langevin Equations (QLEs) that describes the dynamics of our system as, 
\begin{equation}\label{eq:nonli}
\begin{aligned}
\dot{m} &= - (i \Delta_m^\prime + \kappa_m) m - i g_{ma} a - i \tilde{g}_{mb} b_s - i \tilde{g}_{ck} m b_s^\dagger b_s + E + \sqrt{2 \kappa_m} \, m_\mathrm{in}, \\
\dot{a} &= - (i \Delta_a + \gamma_a) a - i g_{ma} m - i \tilde{J}_m b_s + \sqrt{2 \gamma_a} \, a_\mathrm{in}, \\
\dot{b}_s &= - (i \Delta_b + \gamma_b) b_s - i \tilde{g}_{mb} m - i \tilde{J}_m a - i \tilde{g}_{ck} m^\dagger m b_s + \sqrt{2 \gamma_b} \, b_\mathrm{in},
\end{aligned}
\end{equation}
where we have included the dissipations $\kappa_m$, $\gamma_a$, and $\gamma_b$ of the involved modes ($m$, $a$, $b$), together with their related quantum noise $m_{in}$, $a_{in}$, and $b_{in}$, respectively. These noise operators have zero mean values, and are characterized by the following correlation functions,
\begin{equation}\label{eq:6}
\begin{aligned}
\langle b_{s,\mathrm{in}}(t) b_{s,\mathrm{in}}^\dagger(t') \rangle &= (N_s+1)\delta(t-t'), \quad
\langle b_{s,\mathrm{in}}^\dagger(t) b_{s,\mathrm{in}}(t') \rangle &= N_s \delta(t-t'), \\
\langle b_{s,\mathrm{in}}(t) b_{s,\mathrm{in}}(t') \rangle &= M_s \delta(t-t'), \quad
\langle b_{s,\mathrm{in}}^\dagger(t) b_{s,\mathrm{in}}^\dagger(t') \rangle &= M_s^\ast \delta(t-t'),
\end{aligned}
\end{equation}
where the squeezing coefficients are defined as, $N_s= (n_{th}+1)\sinh^2 r + n_{th} \cosh^2 r$,  $M_s= (2n_{th}+1)\sinh r \cosh r$, and the thermal phonon number $n_{th}=\left[\exp\left(\frac{\hbar\omega_b}{k_B\text{T}}\right)-1\right]^{-1}$, with $T$ and $k_B$ the bath temperature and the Boltzmann constant, respectively. The above nonlinear QLEs (in \cref{eq:nonli}) can be linearized by splitting each operator ($\mathcal{O}$) into its mean value ($\langle\mathcal{O}\rangle$) plus an amount of fluctuations ($\delta\mathcal{O}$) around it. This procedure leads to the following mean value equations,
\begin{equation}
\begin{aligned}
0&=(i \tilde{\Delta}_m + \kappa_m) m_s + i g_{ma} a_s + i \tilde{g}_{mb} \beta^* -E \\
0&=(i \Delta_a + \gamma_a) a_s + i g_{ma} m_s + i \tilde{J}_m \beta, \\
0&=(i \tilde{\Delta}_b + \gamma_b) \beta + i \tilde{g}_{mb} m_s^* + i \tilde{J}_m a_s,
\end{aligned}
\end{equation}
with the related set of equations capturing the dynamical fluctuation,
\begin{equation}\label{eq:5}
\begin{aligned}
\delta \dot{m} &= - (i \tilde{\Delta}_m + \kappa_m)  \delta m 
- i g_{ma} \delta a 
-iG_f\delta b_s- i \tilde{G}_{ck}\delta b_s^\dagger  
+ \sqrt{2 \kappa_m} m_\mathrm{in}, \\
\delta \dot{a} &= - (i \Delta_a + \gamma_a)  \delta a 
- i g_{ma} \delta m 
- i \tilde{J}_m \delta b_s 
+ \sqrt{2 \gamma_a}  a_\mathrm{in},\\
\delta \dot{b}_s &= - (i \tilde{\Delta}_b + \gamma_b) \delta b_s 
- i \tilde{J}_m \delta a-iG_f\delta m 
- i \tilde{G}_{ck}\delta m^\dagger  
+\sqrt{2 \gamma_b}  b_\mathrm{in}.
\end{aligned}
\end{equation}
where we have defined the following effective quantities, $G_f=\tilde{G}_{ck}+\tilde{g}_{mb}$, $\tilde{\Delta}_m = \Delta_m^\prime + \tilde{g}_{ck} \beta^2$,  $\tilde{\Delta}_b=\Delta_b+\tilde{g}_{ck}m_s^2$,  and $\tilde{G}_{ck} = \tilde{g}_{ck} m_s \beta$. 
In order to investigate steady-state entanglement in our proposal, we define the following quadrature operators,
\begin{equation}
\begin{aligned}
\delta X_\mathcal{O} &= \frac{\delta \mathcal{O} + \delta \mathcal{O}^\dagger}{\sqrt{2}}, & \delta Y_\mathcal{O} &= \frac{\delta \mathcal{O} - \delta \mathcal{O}^\dagger}{i \sqrt{2}},
\end{aligned}
\end{equation}
together with their related noise quadrature operators,
\begin{equation}
\begin{aligned}
X_{\mathcal{O},\mathrm{in}} &= \frac{\mathcal{O}_\mathrm{in} + \mathcal{O}_\mathrm{in}^\dagger}{\sqrt{2}}, & Y_{\mathcal{O},\mathrm{in}} &= \frac{\mathcal{O}_\mathrm{in} - \mathcal{O}_\mathrm{in}^\dagger}{i \sqrt{2}}.
\end{aligned}
\end{equation}
In terms of the above quadrature operators, our linearized equations displayed in \cref{eq:5} can be written in the following compact form,
\begin{equation}\label{eq:7}
\dot{u}(t)=Au(t)+n(t),
\end{equation} 
where $A$ is a $6\times 6$ matrix which reads,
\begin{align}
A =
\begin{pmatrix}
- {\kappa_m} & \tilde{\Delta}_m & 0 & g_{ma} & 0 &G_f-\tilde{G}_{ck} \\
-  \tilde{\Delta}_m & - {\kappa_m} & -g_{ma} & 0 & - ({G}_{f}+\tilde{G}_{ck}) & 0 \\
0 & g_{ma} & - {\gamma_a} & \Delta_a & 0 & \tilde{J}_m \\
- g_{ma} & 0 & -\Delta_a & - {\gamma_a} & -\tilde{J}_m  & 0 \\
0 & G_f-\tilde{G}_{ck} & 0 & \tilde{J}_m & - {\gamma_b} & \tilde{\Delta}_b \\
-({G}_{f}+\tilde{G}_{ck}) & 0 & - \tilde{J}_m & 0 & - \tilde{\Delta}_b & -{\gamma_b}
\end{pmatrix},
\end{align}
with the column vector $u$ defined as, $u = (\delta X_m, \delta Y_m, \delta X_a, \delta Y_a, \delta X_b, \delta Y_b)^\top$, and the column noise vector $n =(\sqrt{2\kappa_m}X_m^{\text{in}},\sqrt{2\kappa_m}Y_m^{\text{in}},\sqrt{2\gamma_a}X_{a}^{\text{in}},\sqrt{2\gamma_a}Y_{a}^{\text{in}},\sqrt{2\gamma_b}X_b^{\text{in}}, \sqrt{2\gamma_b}Y_b^{\text{in}})^\top$, where ($^\top$) stands for the transposition operation. To quantify steady-state quantum entanglement, our system must be stable, meaning that all the real parts of the eigenvalues of the drift matrix $A$ are negative. This stability conditions is derived through the Routh-Hurwitz criterion \cite{Dejesus1987}. Owing to the Gaussian nature of quantum noises and the linearity of the QLEs, the system can fully be characterised by $6\times6$ covariance matrix (CM) $V$ which, can be obtained by solving the following Lyapunov equation,
\begin{equation}\label{eq:Lya}
AV+VA^{\top}=-D,
\end{equation}
where the CM matrix elements are defined as $V_{lk}=\frac{1}{2}\left\{y_{l}(t)y_{k}(t^{\prime})+y_{k}(t^{\prime})y_{l}(t)\right\}$, and the diffusion matrix is given by,
\begin{equation}
D =\text{diag}[\kappa_m,\kappa_m,\gamma_a,\gamma_a,\gamma_b(2N_s+ 2M_s+1),\gamma_b(2N_s- 2M_s+1)].
\end{equation}
The covariance matrix $V$ obtained from \cref{eq:Lya} is physical only if all eigenvalues of the $6\times6$ drift matrix $A$ have negative real parts. We define $\lambda_{\max}=\max_i\mathrm{Re}[\lambda_i(A)]$ and call the system stable when $\lambda_{\max}<0$.
While the elements of the covariance matrix $V$ can be evaluated, this requires a tedious task. Therefore, the $V_{lk}$ elements will be numerically computed by solving \cref{eq:Lya}, and the extracted elements will be used to quantify entanglement in our system. 

\subsection{Quantification of bipartite entanglement}\label{IIC}
Bipartite entanglement is quantified in our system through  logarithmic negativity $E_{N}$, while the residual minimum contangle $\mathcal{R}_{\tau}^{min}$ is used to quantify genuine tripartite entanglement. The logarithmic negativity is expressed as ~\cite{Plenio2005},
\begin{equation}
E_{N}=\text{max}\left[0,-\text{ln}(2\nu)\right],
\end{equation}
where $\nu\equiv2^{-1/2}\left\{\sum(V)-\left[\sum(V)^{2}-4\text{det}(V)\right]^{1/2}\right\}^{1/2}$ with $\sum(V)=\text{det}(V_{i})+\text{det}(V_{j})-2\text{det}(V_{ij})$. Here, the $V_i$  and $V_j$ are $2 \times 2$ matrices extracted from the $ 6\times6$ covariance matrix $V$, and they capture the dynamics of the $i^{th}$ and $j^{th}$ considered subsystems, while $V_{ij}$ describes quantum correlations between these subsystems. For instance, a submatrix for the considered subsystems labelled as $1$ and $2$ reads,
\begin{equation}
V_{sub12}=
\begin{pmatrix}
V_{1}&V_{12}\\
V_{12}^{\top}&V_{2}\\
\end{pmatrix}.
\end{equation}

\subsection{Quantification of tripartite entanglement}
The tripartite entanglement is quantified by the minimum residual contangle $\mathcal{R}_{\tau}^{min}$ that is defined as ~\cite{Coffman2000}, 

\begin{equation}\label{eq:16}
\mathcal{R}_{\tau}^{min}\equiv \text{min}\left[\mathcal{R}_{\tau}^{m|ab},\mathcal{R}_{\tau}^{a|mb},\mathcal{R}_{\tau}^{b|ma}\right].
\end{equation}
This expression guarantees the invariance of tripartite entanglement under all possible permutations of the modes $\mathcal{R}_{\tau}^{r|st}\equiv \mathcal{C}_{r|st}-\mathcal{C}_{r|s}-\mathcal{C}_{r|t},~(r,s,t=m,a,b)$ satisfying the monogamy of quantum entanglement $\mathcal{R}_{\tau}^{r|st}\geq 0$.
The quantity $\mathcal{C}_{u|v}$ is the contangle of subsystem $u$ and $v$, which can be defined as squared logarithmic negativity and it is a proper entanglement monotone~\cite{Plenio2005},  i.e., $\mathcal{C}_{u|v}=E^2_{n_{u|v}}$, with 
\begin{equation}
E_{n}\equiv \text{max}\left[0,-\ln2\eta\right],
\end{equation}
where 
\begin{equation}
\eta=\text{min}~ \text{eig}[i\Omega_{3}\eta_{6}^\prime].
\end{equation}
In the above expression,  $\Omega_{3}$ and $\eta_{6}^\prime$  are respectively defined as,
\begin{equation}
\Omega_{3}=\bigoplus_{k=1}^3i\sigma_{y},~~ \sigma_{y}=\begin{pmatrix}
0&-i\\i&0
\end{pmatrix}
\end{equation} 
and 
\begin{equation}
\eta^\prime_{6}=P_{r|st}VP_{r|st}~~ \text{for}~~ r,s,t=m,a,b
\end{equation}
where 
\begin{equation}
\begin{aligned}
P_{m|ab}&=\text{diag(1,-1,1,1,1,1)},\quad P_{a|mb}&=\text{diag(1,1,1,-1,1,1)},\quad P_{b|ma}&=\text{diag(1,1,1,1,1,-1)},
\end{aligned}
\end{equation}
are partial transposition matrices, and $V$ is the $6 \times 6$ covariance matrix aforementioned.

\section{Results and discussion}\label{sec:Re}
\begin{figure}[tbh]
	\centering
	\includegraphics[width=15cm]{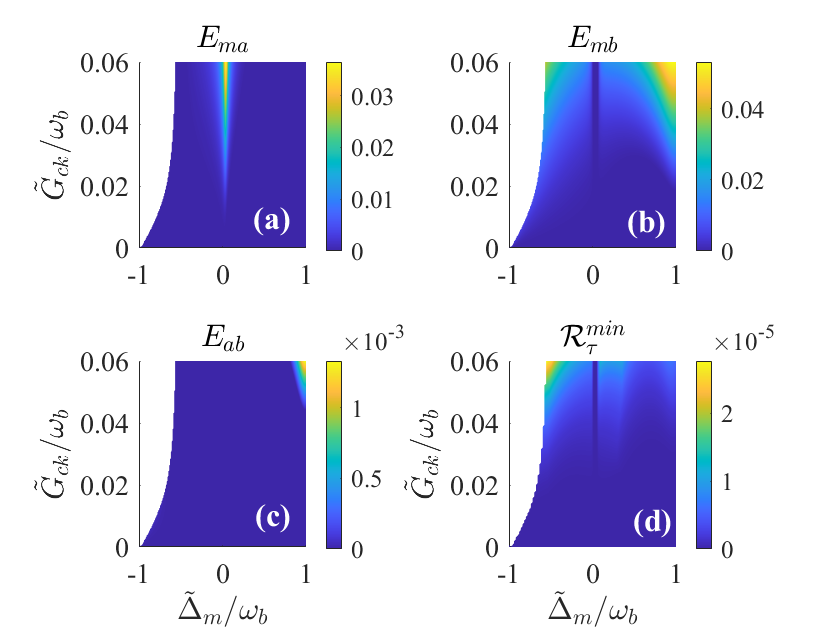}
	\caption{Contour plot of bipartite entanglement (a) for magnon-acoustic modes ($E_{ma}$), (b) for magnon-phonon modes ($E_{mb}$), (c) for acoustic-phonon modes ($E_{ab}$), and (d)  the minimum residual cotangle $\mathcal{R}_\tau^{min}$  as a function of the normalized detunings $\tilde{\Delta}_m$ and effective cross Kerr coupling $\tilde{G}_{ck}$. Common parameters for all subplots are, $\omega_b/2\pi=22.5 $MHz,  $\gamma_b/\omega_b= 2\times 10^{-5}$, $g_{ma}/\omega_b = {0.2}$, $\Delta_a/\omega_b=1$, $g_{mb}/g_{ma}=0.49$, $J_m/\omega_b=0.0$, $\chi/\omega_b=0.0$,  $\kappa_m/g_{ma}=0.02$,   $\gamma_a=0.4\omega_b$, and $T= 10 $mK. In all panels, the white region marks parameters for which the system is unstable ($\lambda_{\max}\ge0$), where the covariance matrix is not defined and no value is plotted. The colored region satisfies $\lambda_{\max}<0$.}
	\label{fig:fig1}
\end{figure}

To carry out our entanglement analysis, we adopt state-of-the-art parameters that are consistent with recent experimental works in optomagnomechanical structures as summarized in \cref{tab:parameters} above. In order to get the optimal value of the cross-Kerr effective coupling strength, we represent the contour plot of entanglement versus effective coupling $\tilde{G}_{ck}$ and the detuning $\tilde{\Delta}_m$ in \cref{fig:fig1}.
%
%
\begin{table*}[t]
	\caption{Parameters used in the numerical calculations. Ratios are given with respect to $\omega_b=2\pi\times22.5\,\mathrm{MHz}$.}
	\label{tab:parameters}
	\renewcommand{\arraystretch}{1.2}
	\begin{tabular}{llll}
		\hline\hline
		Parameter & Ratio & Physical value & Source \\
		\hline
		Mechanical damping rate, $\gamma_b$ & $2\times10^{-5}\omega_b$ & $2\pi\times450\,\mathrm{Hz}$ & Zhang et al.~\cite{Zhang2016} \\
		Magnon-acoustic coupling, $g_{ma}$ & $0.2\omega_b$ & $2\pi\times4.5\,\mathrm{MHz}$ & Xiao-Wen et al.~\cite{Huo2025} \\
		Magnon-phonon coupling, $g_{mb}$ & $0.49\,g_{ma}$ & $\approx2\pi\times2.2\,\mathrm{MHz}$ & Xiao-Wen et al.~\cite{Huo2025} \\
		Phonon hopping rate, $J_m$ & $0.1\omega_b$ & $2\pi\times2.25\,\mathrm{MHz}$ & Zhuo-Zhi et al.~\cite{Zhang2020} \\
		Magnon damping rate, $\kappa_m$ & $0.02\,g_{ma}$ & $2\pi\times90\,\mathrm{kHz}$ & Xiao-Wen et al.~\cite{Huo2025} \\
		Parametric-amplification strength, $\chi$ & $0.5\omega_b$ & $2\pi\times11.25\,\mathrm{MHz}$ & Djorw\'e et al.~\cite{Djorw2024} \\
		Acoustic mode damping rate, $\gamma_a$ & $0.4\omega_b$ & $2\pi\times9\,\mathrm{MHz}$ & D.R.K. Massembele et al.~\cite{MASSEMBELE2025416689} \\
		Magnon detuning, $\Delta_m$ & $\omega_b$ & $2\pi\times22.5\,\mathrm{MHz}$ & This work \\
		Acoustic mode detuning, $\Delta_a$ & $\omega_b$ & $2\pi\times22.5\,\mathrm{MHz}$ & Xiao-Wen et al.~\cite{Huo2025} \\
		Bath temperature, $T$ & -- & $10\,\mathrm{mK}$ & Jie Li et al.~\cite{Li2019} \\
		\hline\hline
	\end{tabular}
\end{table*}
 Indeed, this figure displays bipartite entanglement between magnonic and acoustic modes ($E_{ma}$), magnonic and phononic modes ($E_{mb}$),  acoustic and phononic modes ($E_{ab}$), and  the tripartite entanglement between the involved modes ($\mathcal{R}_\tau^{min}$).  It is worth mentioning here that both  phonon hopping rate and parametric driving  are not accounted.
 As one can see from \cref{fig:fig1}, these generated entanglements, i.e., $E_{ma}$, $E_{mb}$, $E_{ab}$, and $\mathcal{R}_\tau^{min}$ do not peak at the same magnonic detuning  $\tilde{\Delta}_m$. Instead they exhibit a complementary behaviour. When one channel strengthens, another weakens, indicating a transfer or redistribution of entanglement among different bipartite pairs. Moreover, \cref{fig:fig1} shows that our system dwells into red-sideband regime ($\tilde{\Delta}_m/\omega_{b}>0$) for the CMM mode. This regime operation  activates anti-Stokes scattering, which cools the CMM mode and enhances quantum correlations.  For the effective coupling strength $\tilde{G}_{ck}$, our results reveal the existence of a well-defined threshold beyond which entanglement emerges in the system, as illustrated in \cref{fig:fig1}. Specifically, this threshold is found to be $\tilde{G}_{ck}\approx 0.04\omega_{b}$ for, $E_{ab}$ (see \cref{fig:fig1}c), while a lower value $\tilde{G}_{ck}\approx 0.02\omega_{b}$ is sufficient for the onset of  $E_{ma}$, $E_{mb}$, and $R^{min}_{\tau}$ (see \cref{fig:fig1} a, b, d). These observations underscore the crucial role of the cross-Kerr nonlinearity as a resource for generating and controlling quantum entanglement. In our system, the cross-Kerr interaction acts directly between the magnonic mode and the CMM mode. This direct coupling explains why the entanglement $E_{mb}$, exhibits a lower threshold compared to other bipartite correlations, which rely on indirect interaction pathways.  Importantly, the results demonstrate that the effective cross-Kerr interaction alone can mediate entanglement among the modes, even in the absence of direct phonon hopping $(J_m=0)$ and without mechanical parametric driving $(\chi=0)$. This highlights the ability of cross-Kerr nonlinearity to act as an effective indirect coupling mechanism, enabling entanglement generation in otherwise uncoupled subsystems.
\begin{figure}[tbh]
	\centering
	\includegraphics[width=15cm]{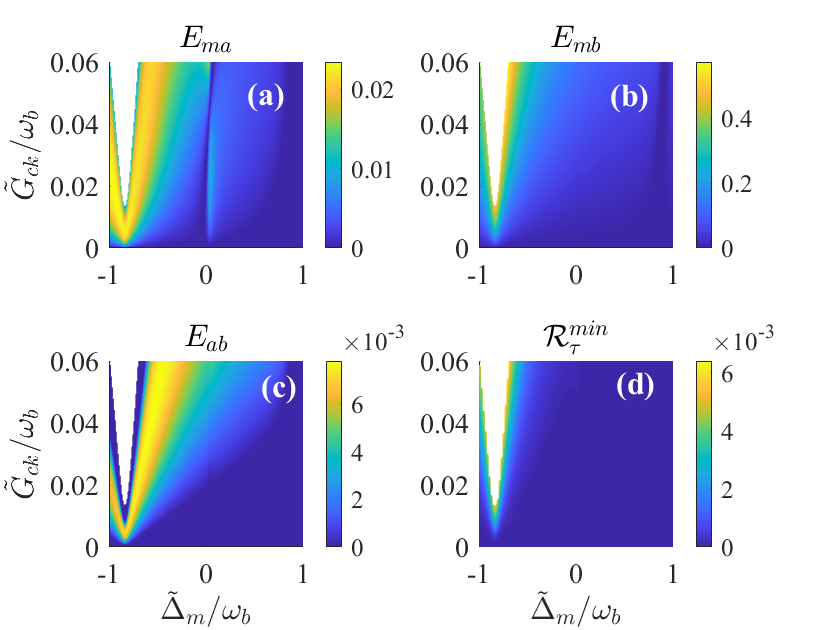}
	\caption{Contour plot of bipartite entanglement (a) for magnon-acoustic modes ($E_{ma}$), (b) for magnon-phonon modes ($E_{mb}$), (c) for acoustic-phonon modes ($E_{ab}$), and (d)  the minimum residual cotangle $\mathcal{R}_\tau^{min}$ as a function of the normalized detunings $\tilde{\Delta}_m$ and effective coupling $\tilde{G}_{ck}$. Common parameters for all subplots are, $\omega_b/2\pi=22.5 $MHz,  $\gamma_b/\omega_b= 2\times 10^{-5}$, $g_{ma}/\omega_b = {0.2}$, $g_{mb}/g_{ma}=0.49$, $J_m/\omega_b=0.1$, $\chi/\omega_b=0.5$,  $\kappa_m/g_{ma}=0.02$, $\Delta_a/\omega_b=1$,  $\gamma_a=0.4\omega_b$, and $T= 10 $mK. In all panels, the white region marks parameters for which the system is unstable ($\lambda_{\max}\ge0$), where the covariance matrix is not defined and no value is plotted. The colored region satisfies $\lambda_{\max}<0$.}
	\label{fig:fig2}
\end{figure}

Taking into account the phonon hopping rate $(J_m \neq 0)$ and the parametric driving amplitude ($\chi\neq0$), \cref{fig:fig2} illustrates that a very low threshold value of $\tilde{G}_{ck}$ is required for the generation of entanglement. This reveals that the phonon hopping rate together with the parametric driving amplification are crucial parameters to tune in order to efficiently generate and enhance entanglement in our proposal. Furthermore, their combined effect extends the regime of entanglement generation even into the modest cross-Kerr  interaction limit.
 Based on the results in \cref{fig:fig1,fig:fig2}, we treat the effective coupling
 	$\tilde G_{ck}$ and the effective magnon detuning $\tilde\Delta_m$ as independent
 	parameters, fixed to $\tilde G_{ck}/\omega_b=0.05$ and $\tilde\Delta_m/\omega_b=0.5$
 	(or $1$), because the linearized dynamics depends on the Kerr coupling only through these
 	combinations. Taking $m_s$ and $\beta$ real, they are related to the bare coupling by
 	$
 	\tilde G_{ck} = g_{ck}\cosh(2r)\, m_s\beta$, and
 	$\tilde\Delta_m = \Delta_m + g_{ck}\sinh^2 r + g_{ck}\cosh(2r)\,\beta^2,
 	$ with $\tilde\Delta_b=\Delta_b+g_{ck}\cosh(2r)\,m_s^2$ for the mechanical mode.
 	For $\chi/\tilde\omega_b=0.5$ the squeezing parameter is $r=\tfrac14\ln3\approx0.27$, so
 	$\tilde g_{ck}=g_{ck}\cosh 2r\approx1.15\,g_{ck}$ and $g_{ck}\sinh^2 r\approx0.08\,g_{ck}$:
 	squeezing modifies the Kerr terms only slightly, and the large ratio $\tilde G_{ck}/g_{ck}$
 	comes from the steady-state amplitudes, $\tilde G_{ck}/g_{ck}\approx1.15\,m_s\beta$. The bare coupling $g_{ck}$ is weak, whereas $\tilde G_{ck}/\omega_b=0.05$ is about
 	twelve times $\kappa_m$ and of the same order as $J_m$, i.e.\ strong with respect to the
 	dissipation rates. Entanglement therefore arises from the drive-enhanced effective coupling, not from a strong bare coupling.
\begin{figure}[tbh]
	\centering
	\includegraphics[width=15cm]{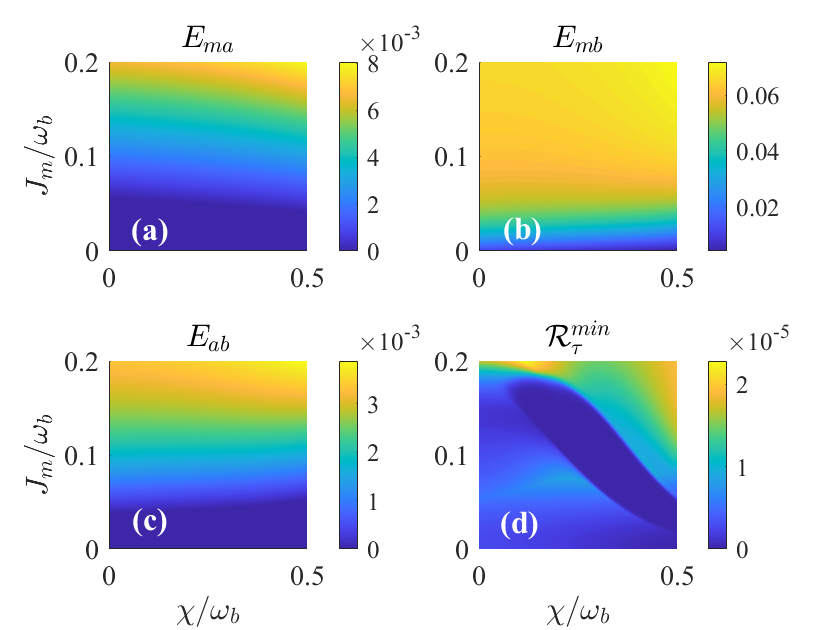}
	\caption{Contour plot of (a) bipartite entanglement for magnon-acoustic modes ($E_{ma}$), when $\tilde{\Delta}_m/\omega_b=0.5$, (b)  bipartite entanglement for magnon-phonon modes ($E_{mb}$), when $\tilde{\Delta}_m/\omega_b=0.5$, (c)  bipartite entanglement for acoustic-phonon modes ($E_{ab}$), (d)  the minimum residual cotangle $\mathcal{R}_\tau^{min}$, when $\tilde{\Delta}_m/\omega_b=1$, as a function of the MPA strength, $\chi$ and phonon hopping, $J_m$. Common parameters for all subplots are, $\omega_b/2\pi=22.5 $MHz,  $\gamma_b/\omega_b= 2\times 10^{-5}$, $g_{ma}/\omega_b = {0.2}$, $g_{mb}/g_{ma}=0.49$, $\tilde{G}_{ck}/\omega_b=0.05$  $\kappa_m/g_{ma}=0.02$, $\Delta_a/\omega_b=1$, $\tilde{\Delta}_m/\omega_b=0.5$,  $\gamma_a=0.4\omega_b$, and $T= 10 $mK.}
	\label{fig:fig4}
\end{figure}
In \cref{fig:fig4}, we show contour plots of entanglement as a function of the mechanical coupling $J_m$ and the parametric driving amplitude $\chi$, in the presence of the cross-Kerr interaction ($\tilde{G}_{ck} \neq 0$). It can be seen that the threshold value of $J_m$ required for the generation of entanglement is lower for $E_{mb}$, and $E_{ab}$, and $R_{\tau}^{min}$, i.e., $J_m \lesssim 0.04\omega_{b}$, while a higher value is needed for $E_{ma}$, with $J_m \approx 0.07\omega_{b}$. This shows that the cross-Kerr nonlinearity, phonon hopping and MPA help to strengthen quantum correlations, although the required mechanical coupling depends on the type of the considered entanglement. The strong enhancement of magnon-phonon correlation ($E_{mb}\approx 0.06$) is due to the fact that the cross-Kerr interaction directly couples these two modes, leading to an efficient generation of phase-dependent correlations without requiring intermediate coupling channels. As a result, entanglement between these two modes is more robust ($E_{mb}\approx 0.06$) compared to other bipartitions in the system. Moreover, we also observe the lowest threshold of $J_m$ that is required for the generation of $E_{mb}$, which relies on the fact that cross-Kerr coupling fosters interaction between these two modes.
\begin{figure}[tbh]
	\centering
	\includegraphics[width=15cm]{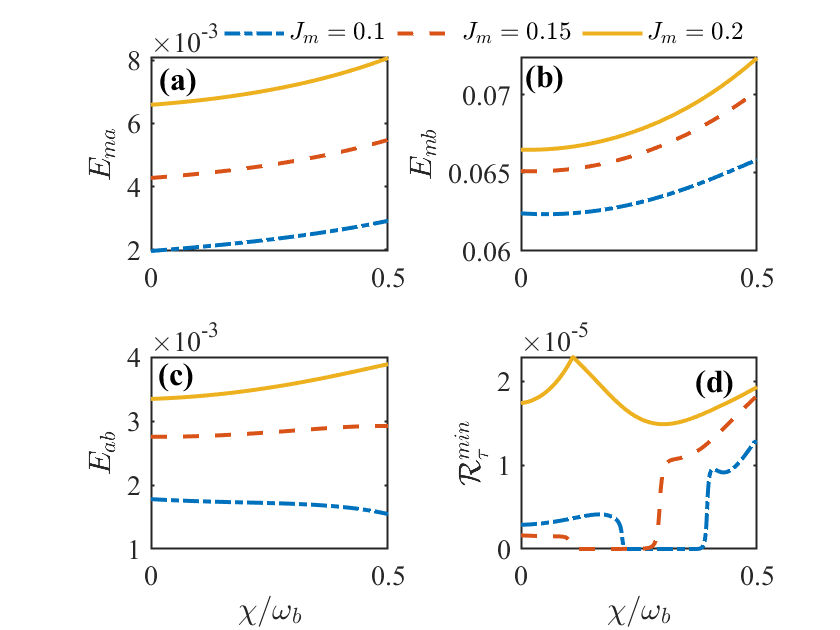}
	\caption{The entanglement degree (a)  $E_{ma}$, when $\tilde{\Delta}_m/\omega_b=0.5$, (b) $E_{mb}$, when $\tilde{\Delta}_m/\omega_b=0.5$, (c) $E_{ab}$, when $\tilde{\Delta}_m/\omega_b=0.5$, (d) $\mathcal{R}_\tau^{min}$, when $\tilde{\Delta}_m/\omega_b=1$, versus $\chi$ for different values of $J_m$. Common parameters for all subplots are, $\omega_b/2\pi=22.5 $MHz,  $\gamma_b/\omega_b= 2\times 10^{-5}$, $g_{ma}/\omega_b = {0.2}$, $g_{mb}/g_{ma}=0.49$,  $\kappa_m/g_{ma}=0.02$, $\Delta_a/\omega_b=1$, $\tilde{G}_{ck}/\omega_b=0.05$,  $\gamma_a=0.4\omega_b$, and $T= 10 $mK.}
	\label{fig:fig31}
\end{figure}
\begin{figure}[tbh]
	\centering
	\includegraphics[width=15cm]{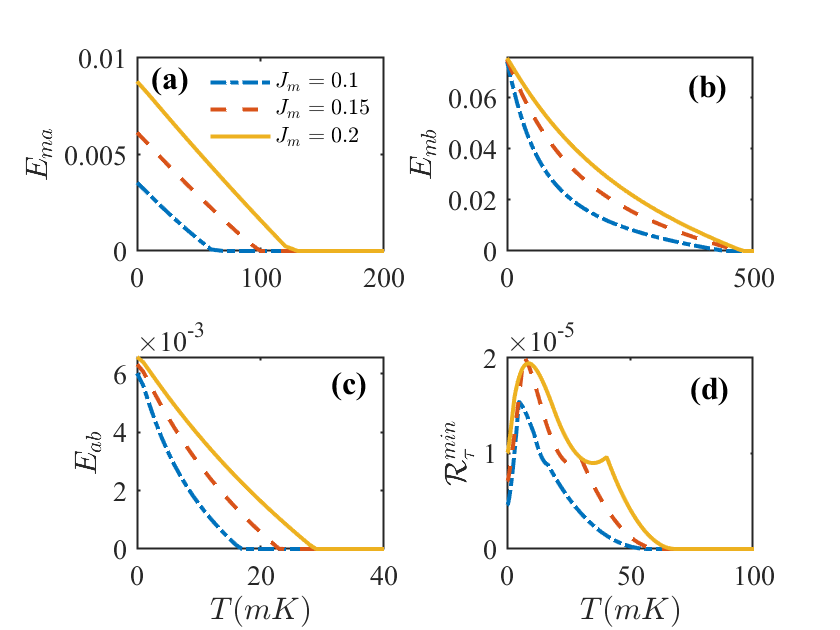}
	\caption{Plot of (a) bipartite entanglement for magnon-acoustic modes ($E_{ma}$), when $\tilde{\Delta}_m/\omega_b=0.5$, (b)  bipartite entanglement for magnon-phonon modes ($E_{mb}$), when $\tilde{\Delta}_m/\omega_b=0.5$, (c)  bipartite entanglement for acoustic-phonon modes ($E_{ab}$), when $\tilde{\Delta}_m/\omega_b=0.5$, (d)  the minimum residual cotangle $\mathcal{R}_\tau^{min}$, when $\tilde{\Delta}_m/\omega_b=1$, as a function of $T$ for different values of $J_m$. Common parameters for all subplots are, $\omega_b/2\pi=22.5 $MHz,  $\gamma_b/\omega_b= 2\times 10^{-5}$, $g_{ma}/\omega_b = {0.2}$, $g_{mb}/g_{ma}=0.49$, $\tilde{G}_{ck}/\omega_b=0.05$,  $\kappa_m/g_{ma}=0.02$, $\Delta_a/\omega_b=1$, $\chi/\omega_b=0.5$, and  $\gamma_a=0.4\omega_b$.}
	\label{fig:fig5}
\end{figure}

In \cref{fig:fig31}, entanglement is plotted over parametric driving amplitude $\chi$ for different values of the phonon hopping rate $J_m$ when ($\tilde{G}_{ck} \neq 0$). In fact, this figure is extracted from \cref{fig:fig4}. It can be seen from \cref{fig:fig31} that the strength of entanglement  grows with the phonon hopping rate $J_m$. In particular, increasing the mechanical coupling leads to a progressive improvement of quantum correlations across the system, ultimately reaching an optimal value at $J_m=0.15\omega_{b}$. This behaviour highlights the crucial role of phonon hopping in strengthening and stabilizing entanglement in the hybrid system.

Another important aspect of the analysis concerns the robustness of entanglement against thermal noise. This investigation provides insight into how effectively the generated entanglement persists when the system interacts with its surrounding thermal environment. For this purpose, \cref{fig:fig5} displays the plot of entanglement versus temperature ($T$) for different values of the phonon hopping rate $J_m$. It is evident from \cref{fig:fig5} that the entanglement gradually vanishes as the temperature increases. In addition, it can be observed that the entanglement becomes more robust against thermal noise with increasing phonon hopping rate $J_m$, revealing an enhanced resilience of the system under stronger mechanical coupling.

\section{Quantum complete synchronization}\label{sec:purity}
Here we investigate quantum complete synchronization which is a measure that describes synchronization solely at the level of quantum fluctuations and is applicable even in the absence of a limit cycle. Furthermore we explore the emergence of quantum complete synchronization between an acoustic phononic mode and CMM phonon mode. While the concept of synchronization traces its origins to Huygens's observation of coupled pendulum clocks, the present work focuses on synchronization in the quantum domain. Unlike classical synchronization, quantum synchronization arises from the non-commutative properties of the canonical position and momentum operators describing the acoustic phononic mode and CMM phonon mode. Based on the formalism proposed by Mari et al., in \cite{PhysRevLett.111.103605}, the criterion for quantum complete synchronization is given by 
\begin{equation}
S_q=\frac{1}{\langle\delta X_-^2+\delta Y_-^2\rangle},
\end{equation}
here, $\delta X_-=(\delta X_{a}-\delta X_{b})/\sqrt{2}$ and $\delta Y_-=(\delta Y_{a}-\delta Y_{b})/\sqrt{2}$. With the help of the covariance matrix $V$ defined above, complete quantum synchronization measure $(S_q)$ can be written as 
\begin{equation}
S_q=\frac{1}{\frac{1}{2}(V_{33}+V_{44}+V_{55}+V_{66}-2V_{35}-2V_{46})},
\end{equation}
where $0<S_q\leq1$. Perfect quantum synchronization is attained for $S_q=1$, while smaller values of $S_q$ indicate weaker synchronization.
\begin{figure}[tbh]
	\centering
	\includegraphics[width=15cm]{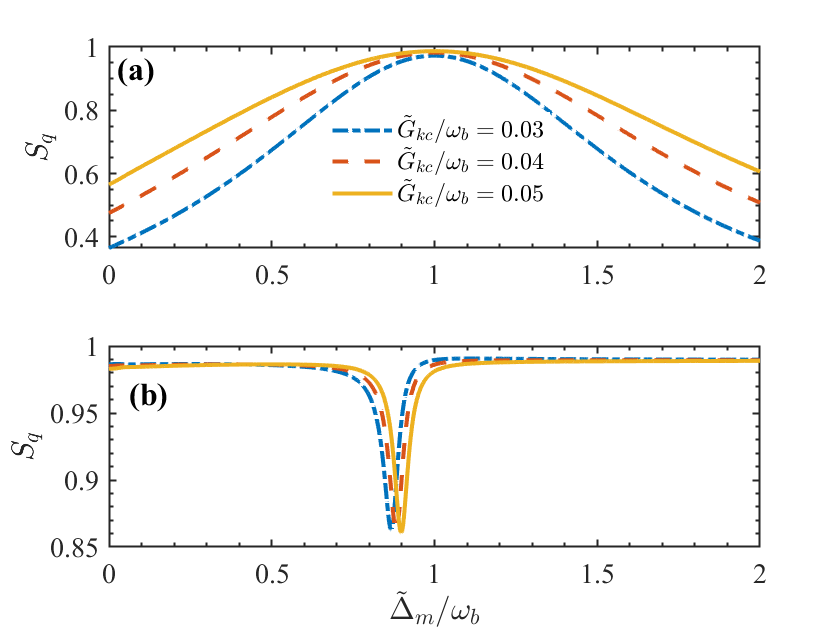}
	\caption{Plot of (a) quantum synchronization measure ($S_q$) as a function of the normalized detunings $\tilde{\Delta}_m$ when $J_m/\omega_b=0.0$, $\chi/\omega_b=0.0$, (b)  quantum synchronization measure ($S_q$) as a function of the normalized detuning $\tilde{\Delta}_m$ when $J_m/\omega_b=0.1$, $\chi/\omega_b=0.5$. Common parameters for all subplots are, $\omega_b/2\pi=22.5 $MHz,  $\gamma_b/\omega_b= 2\times 10^{-5}$, $g_{ma}/\omega_b = {0.2}$, $g_{mb}/g_{ma}=0.49$, $\kappa_m/g_{ma}=0.02$, $\Delta_a/\omega_b=1$, $\tilde{\Delta}_m/\omega_b=0.5$, and  $\gamma_a=0.4\omega_b$.}
	\label{fig:fig6}
\end{figure}

\Cref{fig:fig6} illustrates the quantum synchronization measure $S_q$ as a function of the normalized detuning $\tilde{\Delta}_m$. \Cref{fig:fig6} indicates that acoustic phononic mode and CMM phonon mode always maintain complete quantum synchronization, i.e., $S_q\approx1$ when $\tilde{G}_{ck}/\omega_b\gtrsim0.05$, $J_m/\omega_b=0.0$, $\chi/\omega_b=0.0$ and $\tilde{\Delta}_m\approx\omega_b$. The underlying mechanism responsible for the observed enhancement of quantum synchronization can be attributed to the effective coupling established between the acoustic phononic mode and CMM phonon mode through the magnon mediator. In the absence of the direct phonon hopping ($J_m=0$), the acoustic phononic mode and CMM phonon mode are not completely decoupled, since both interact with the magnon mode. The magnon therefore provide an indirect channel for the exchange of quantum fluctuations, leading to correlations between the mechanical modes and enabling quantum synchronization. Moreover, the cross-Kerr interaction modifies the effective magnon and CMM detunings and introduces an additional contribution to the linearised magnon-CMM coupling through $\tilde{G}_{ck}$. This effective coupling improved synchronization between the acoustic phononic mode and CMM phonon mode. 
\begin{figure}[tbh]
	\centering
	\includegraphics[width=15cm]{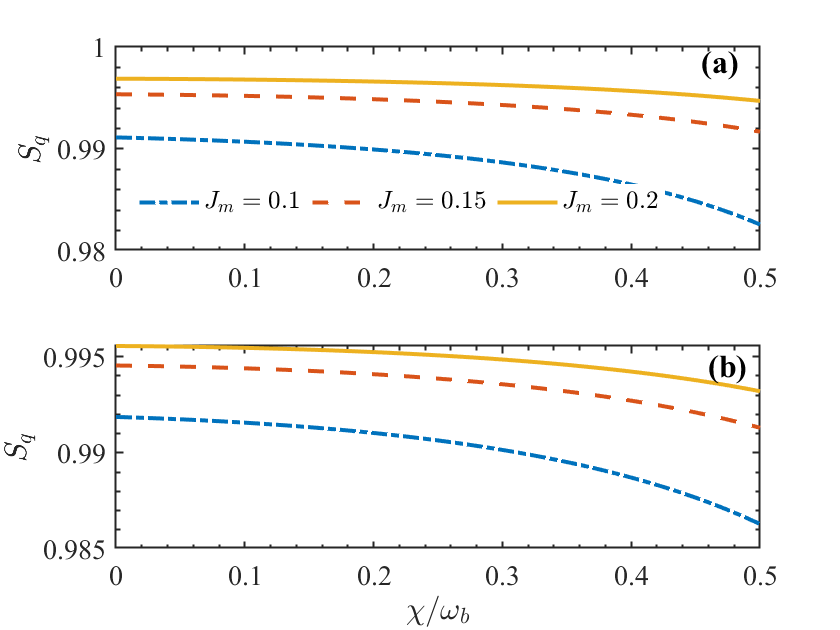}
	\caption{Plot of (a) quantum synchronization measure ($S_q$) as a function of, $\chi$ for different values of phonon hopping rate $J_m$ when $\tilde{G}_{ck}/\omega_b=0.0$ (b)  quantum synchronization measure ($S_q$) as a function, $\chi$ for different values of phonon hopping rate $J_m$ when $\tilde{G}_{ck}/\omega_b=0.05$. Common parameters for all subplots are, $\omega_b/2\pi=22.5 $MHz,  $\gamma_b/\omega_b= 2\times 10^{-5}$, $g_{ma}/\omega_b = {0.2}$, $g_{mb}/g_{ma}=0.49$,  $\kappa_m/g_{ma}=0.02$, $\Delta_a/\omega_b=1$, $\tilde{\Delta}_m/\omega_b=0.5$, and  $\gamma_a=0.4\omega_b$.}
	\label{fig:fig7}
\end{figure}

\Cref{fig:fig7} shows the quantum synchronization measure $S_q$ as a function of $\chi$ for different values of phonon-hopping $J_m$. As one can see clearly from this figure, for low values of $\chi$, the synchronization between the acoustic mode and CCM mode is mainly governed by the direct phonon-hopping interaction and the magnon-mediated coupling, resulting in stronger phase locking and reduced synchronization error. As $J_m$ increases, the coherent exchange of quantum fluctuations between the two mechanical modes is enhanced, leading to an increase in $S_q$.  However, increasing the parametric amplification strength $\chi$, introduces phase-sensitive noise and quadrature anti-squeezing into the CMM mode. Consequently, the complete synchronization error increases, causing a reduction in the quantum synchronization measure. This indicates that, although mechanical parametric amplification can enhance certain quantum correlations, it is detrimental to quantum synchronization.
\begin{figure}[tbh]
	\centering
	\includegraphics[width=10cm]{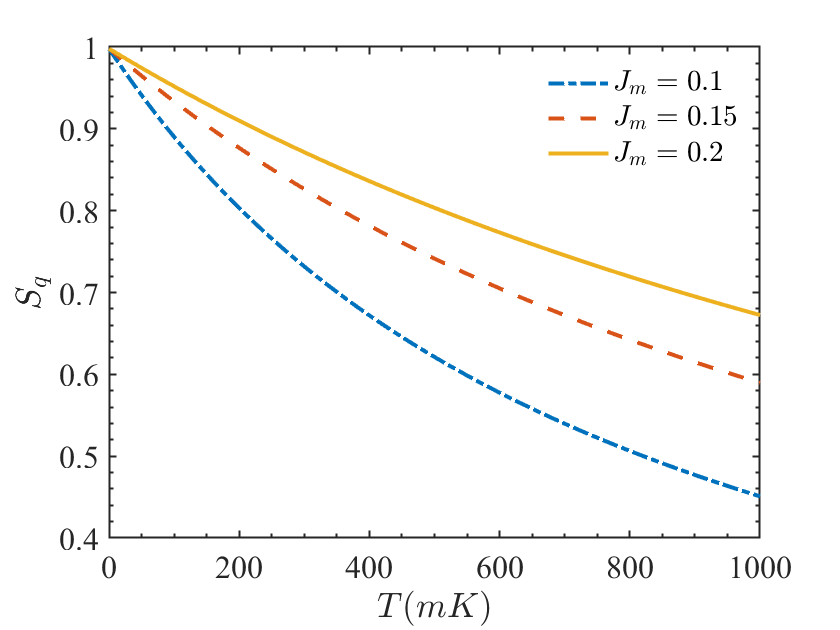}
	\caption{Plot of quantum synchronization measure ($S_q$) as a funcction of temperature. Common parameters for all subplots are, $\omega_b/2\pi=22.5 $MHz,  $\gamma_b/\omega_b= 2\times 10^{-5}$, $g_{ma}/\omega_b = {0.2}$, $g_{mb}/g_{ma}=0.49$, $\tilde{G}_{ck}/\omega_b=0.05$  $\kappa_m/g_{ma}=0.02$, $\Delta_a/\omega_b=1$, $\tilde{\Delta}_m/\omega_b=0.5$, $\chi/\omega_b=0.3$ and  $\gamma_a=0.4\omega_b$.}
	\label{fig:fig8}
\end{figure}

In \Cref{fig:fig8}, we illustrate the impact of ambient temperature on quantum complete synchronization.
As the ambient temperature increases, the quantum synchronization monotonically degrades due to enhanced thermal phonon occupancy. This thermal injection increases the mechanical quadrature noise variances and suppress the intermode quantum coherence between the acoustic mode and CMM mode. These results demonstrate the inherent sensitivity of quantum synchronization to thermal decoherence, while also highlighting a low-temperature regime where $S_q$ remains robust against weak thermal fluctuations.
\section{Experimental feasibility}\label{sec:feasiy}
The scheme can be implemented with a YIG sphere of diameter about $125\,\mu\mathrm{m}$
	placed in a uniform bias field $B_0$, which sets the Kittel-mode magnon frequency
	$\omega_m=\gamma B_0$, where $\gamma$ is the gyromagnetic ratio. The sphere supports a
	magnon mode together with two mechanical modes, the deformation mode and the
	center-of-mass motion (CMM), both demonstrated in magnomechanical	experiments~\cite{Zhang2016}. The deformation-mode frequency scales inversely with the
	sphere diameter: a $250\,\mu\mathrm{m}$ sphere gives $\omega_b/2\pi=11.42\,\mathrm{MHz}$~\cite{Zhang2016},
	so $\omega_b/2\pi=22.5\,\mathrm{MHz}$ corresponds to roughly half that diameter. The magnon
	mode is driven by a weak gradient field $H_d$, following~\cite{Huo2025}.

 \emph{Phonon hopping.} The value $J_m/\omega_b=0.1$ corresponds to
	$J_m/2\pi\approx2.3\,\mathrm{MHz}$. The coupling between the two mechanical modes can be
	generated parametrically: modulating the stress (spring constant) of the resonator at the
	difference frequency of the two modes yields an effective beam-splitter interaction,
	$J_m(a^\dagger b+ab^\dagger)$, whose strength is set by the modulation amplitude and can
	therefore be tuned~\cite{Okamoto2013}. Phonon hopping rates of several MHz,
	exceeding the intrinsic mechanical dissipation, have been demonstrated in coupled
	mechanical resonators~\cite{Okamoto2013,Zhang2020}.
	 It satisfies $J_m\sinh r\ll\Delta_a+\Delta_b$, which
	justifies the rotating-wave approximation of \cref{sec:Appen}.
	
	\emph{Parametric amplification.} The value $\chi/\omega_b=0.5$ is realized by modulating the spring constant of the CMM
	at frequency $2\omega_d$ with phase $\phi$, a technique used for parametric squeezing of mechanical modes~\cite{Rugar1991,Pontin2014}.
 	
	\emph{Cross-Kerr coupling.} Kerr-type nonlinearities have been observed in cavity magnomechanics~\cite{Shen}. Our calculations use the effective coupling
	$\tilde G_{ck}/\omega_b=0.05$, about twelve times $\kappa_m$ and of the same order as $J_m$. It follows from the bare coupling $g_{ck}$ through $\tilde G_{ck}=\tilde g_{ck}\,m_s\beta$, where
	$m_s$ and $\beta$ are the steady-state magnon and phonon amplitudes.
\section{Conclusion}\label{sec:concl}
In this work, we have investigated the generation of bipartite and tripartite entanglement, together with quantum complete synchronization, in a hybrid magnomechanical system composed of magnon, acoustic, and center-of-mass mechanical modes. Our results show that, in the absence of mechanical coupling ($J_m$), and parametric driving ($\chi$), the generation of entanglement requires a relatively strong effective cross-Kerr interaction. However, when phonon hopping and parametric amplification are included, quantum entanglement can be achieved even in the weak-coupling regime of the effective cross-Kerr nonlinearity. In particular, phonon hopping plays a key role in redistributing quantum correlations across the system, enabling efficient generation of both bipartite and tripartite entanglement.

 Furthermore, we have demonstrated that the coupled acoustic mode and CMM mode exhibit quantum complete synchronization arising from the interplay between the direct phonon-hopping interaction and the indirect magnon-mediated coupling. We have shown that the cross-Kerr nonlinearity provides an effective means of controlling the synchronization dynamics, whereas the MPA and thermal fluctuations strongly influence the degree of synchronization. These results establish the proposed hybrid magnomechanical system as a versatile platform for the simultaneous engineering of entanglement and quantum synchronization. Our findings highlight that the joint action of phonon hopping, parametric driving, and cross-Kerr nonlinearity provides a versatile platform for engineering robust quantum entanglement in hybrid magnomechanical systems. These results may have potential applications in quantum information processing, quantum sensing, and the development of hybrid quantum networks.

\section*{Acknowledgments}
P.D. acknowledges the Iso-Lomso Fellowship at Stellenbosch Institute for Advanced Study (STIAS), Wallenberg Research Centre at Stellenbosch University, Stellenbosch 7600, South Africa, and The Institute for Advanced Study, Wissenschaftskolleg zu Berlin, Wallotstrasse 19, 14193 Berlin, Germany. This work was supported by Princess Nourah
bint Abdulrahman University Researchers Supporting Project number (PNURSP2026R893), Princess Nourah bint Abdulrahman University, Riyadh, Saudi Arabia. The authors are thankful to the Deanship of Graduate Studies and Scientific Research at University of Bisha for supporting this work through the Fast-Track Research Support Program
\section*{Author Contributions}
E.K.B. and P.D. conceptualized the work and carried out the simulations and analysis. A.N.A.-A and H.A. participated in all the discussions and provided useful methodology and suggestions for the final version of the manuscript. P.D. and A.-H. A.-A. participated in the discussions and supervised the work. All authors participated equally in the writing, discussions, and the preparation of the final version of the manuscript.

\section*{Competing Interests} 
All authors declare no competing interests.

\section*{Data Availability}
Relevant data are included in the manuscript and supporting information. Supplementary data are available upon reasonable request.
\appendix
\section{Derivation of the effective Hamiltonian}\label{sec:Appen}
In this appendix, we present the derivation of the effective Hamiltonian from the full Hamiltonian of the system, given in \Cref{eq:hamil}, which reads
\begin{equation}\label{eq:THam}
\begin{aligned}
\mathcal{H}&= \Delta_m\, m^\dagger m + \Delta_a\, a^\dagger a + \tilde{\omega}_b\, b^\dagger b  + g_{ma} (m a^\dagger + m^\dagger a) + g_{mb} (m b + m^\dagger b^\dagger) + J_m (a^\dagger b + a b^\dagger)+ g_{ck} m^\dagger m b^\dagger b + \frac{\chi}{2}(b^{\dagger 2} +b^2) + iE(m^\dagger - m).
\end{aligned}
\end{equation}
\Cref{eq:THam} can be diagonalize by introducing the squeezing transformation
$S(r) = \exp\left[\frac{r}{2}\left(b^2 - b^{\dagger 2}\right)\right]$,
where $r$ is the squeezing parameter. Under this Bogoliubov transformation, the annihilation operator $b$ transforms into $b_s$ according to
\begin{equation}
b_s = S b S^\dagger = b\cosh r + b^\dagger \sinh r,
\qquad
b = b_s\cosh r - b_s^\dagger \sinh r .
\end{equation}
Using $[b_s,b_s^\dagger]=1$.
The mechanical part of \Cref{eq:THam} is
\begin{equation}
H_{m} = \tilde{\omega}_b b^\dagger b
+ \frac{\chi}{2}\left(b^2 + b^{\dagger 2}\right),
\label{eq:H_mech_initial}
\end{equation}
and is transform as
\begin{align}
b^\dagger b
&= (cosh^2r + \sinh^2 r)\, b_s^\dagger b_s + \sinh^2 r
- \cosh r \sinh r\left(b_s^2 + b_s^{\dagger 2}\right) \nonumber\\
&= \cosh(2r)\, b_s^\dagger b_s + \sinh^2 r
- \frac{1}{2}\sinh(2r)\left(b_s^2 + b_s^{\dagger 2}\right).
\label{eq:bdaggerb_transform}
\end{align}
and 
\begin{equation}
\begin{aligned}
\frac{\chi}{2} (b^{\dagger 2} + b^2) &= \frac{\chi}{2} \Big[ (b_s^{\dagger 2} + b_s^2)(\cosh^2 r + \sinh^2 r)- 4 b_s^\dagger b_s \cosh r \sinh r - 2 \cosh r \sinh r \Big]
\end{aligned}
\end{equation}
The anomalous terms vanish when
$
\chi\cosh(2r) = \omega_b\sinh(2r),
$
consequently,
\begin{equation}
 r = \frac{1}{4} \ln \frac{\tilde{\omega}_b + \chi}{\tilde{\omega}_b - \chi}.
\end{equation}
The effective mechanical frequency is the coefficient of \(b_{s}^\dagger b_{s}\),
$
\Delta_b = \tilde{\omega}_b \cosh 2r - \chi \sinh 2r.
$Using the hyperbolic identities:
\begin{equation}
\cosh 2r = \frac{\tilde{\omega}_b}{\sqrt{\tilde{\omega}_b^2 - \chi^2}}, \quad 
\sinh 2r = \frac{\chi}{\sqrt{\tilde{\omega}_b^2 - \chi^2}}.
\end{equation}
then, 
\begin{equation}
\Delta_b = \sqrt{\tilde{\omega}_b^2 - \chi^2}.
\end{equation}
The hopping interaction is 
\begin{eqnarray}
J_m (a^\dagger b + a b^\dagger) &= J_m \Big[ a^\dagger (b_s \cosh r - b_s^\dagger \sinh r) + a (b_s^\dagger \cosh r - b_s \sinh r) \Big]\\
&=J_m \cosh r\left(a^\dagger b_s + a b_s^\dagger\right)
- J_m \sinh r\left(a^\dagger b_s^\dagger + a b_s\right).
\end{eqnarray} 
Under the RWA, 
$
J_m (a^\dagger b + a b^\dagger) = \tilde{J}_m\left(a^\dagger b_s + a b_s^\dagger\right),
$
where
$
\tilde{J}_m = J_m\cosh r.$
  The magnon-phonon interaction is
transform as $$g_{mb} (m b + m^\dagger b^\dagger) = g_{mb} \cosh r (m b_s + m^\dagger b_s^\dagger) - g_{mb} \sinh r (m^\dagger b_s + m b_s^\dagger),$$ the term $-g_{mb} \sinh r (m^\dagger b_s + m b_s^\dagger)$ rotates at $\Delta_m' - \Delta_b$, so it is slow and is kept. The term $g_{mb} \cosh r (m b_s + m^\dagger b_s^\dagger)$ rotates at $\Delta_m' + \Delta_b \approx 2\tilde{\omega}_b$, so it is fast and is dropped and then, 
\begin{equation}
g_{mb}\left(m b^\dagger + m^\dagger b\right) = \tilde{g}_{mb}\left(m b_s^\dagger + m^\dagger b_s\right)
\end{equation}
where
$ \tilde{g}_{mb}=-g_{mb}\sinh r$. For the cross-Kerr interaction, the final term contains the anomalous phonon operators and is neglected. We
 therefore obtain
 \begin{equation}\label{eq:Kerr}
 H_{ck}^{\rm RWA}
 	= \tilde{g}_{ck}\, m^\dagger m\, b_s^\dagger b_s
 	+  g_{ck}\sinh^2 r\, m^\dagger m,
 \end{equation}
 where
 $\tilde{g}_{ck} = g_{ck}\cosh(2r).
$
 The second term in \Cref{eq:Kerr} renormalizes the magnon detuning.
 Thus, we define
$ \Delta_m^\prime = \Delta_m + g_{ck}\sinh^2 r$. Combining all the expression above, we can write the transformed Hamiltonian as
\begin{equation}
\begin{aligned}
\tilde{\mathcal{H}} &= \Delta_m^\prime m^\dagger m + \Delta_a a^\dagger a + \Delta_b b_s^\dagger b_s + g_{ma} (m a^\dagger + m^\dagger a) + \tilde{g}_{mb} (m^\dagger b_s + m b_s^\dagger) +  \tilde{J}_m (a^\dagger b_s + a b_s^\dagger)  + \tilde{g}_{ck}, m^\dagger m b_s^\dagger b_s + i E (m^\dagger - m),
\end{aligned}
\end{equation}
\section{Derivation of the diffusion matrix}
The diffusion matrix $D$ is defined via the symmetric noise correlation function
\begin{equation}
\tfrac12 \langle n_i(t) n_j(t') + n_j(t') n_i(t) \rangle = D_{ij} \, \delta(t-t'),
\end{equation}
where $\mathbf{n}$ is the vector of quadrature noise operators. Because the three baths are independent, $D$ decomposes into a direct sum of three $2 \times 2$ blocks. For the magnon and acoustic modes (with thermal occupations $N_m$ and $N_a$), the corresponding blocks are given by $\Gamma_j (2N_j + 1) \, \mathbf{1}_2$, where $\Gamma_j \in \{\kappa_m, \gamma_a\}$.
The mechanical mode is damped by a thermal bath with mean occupation $n_{th}=\left[\exp\left(\frac{\hbar\omega_b}{k_B\text{T}}\right)-1\right]^{-1}$. In the squeezed frame, the input noise operators $b_{s,\mathrm{in}}$ satisfy the correlations
\begin{equation}\label{eq:67}
\begin{aligned}
\langle b_{s,\mathrm{in}}(t) b_{s,\mathrm{in}}^\dagger(t') \rangle &= (N_s + 1) \, \delta(t-t'), \qquad
\langle b_{s,\mathrm{in}}^\dagger(t) b_{s,\mathrm{in}}(t') \rangle &= N_s \, \delta(t-t'),
\end{aligned}
\end{equation}
where $N_s = (n_{\text{th}} + 1)\sinh^2 r + n_{\text{th}} \cosh^2 r$ and $M_s = (2n_{\text{th}} + 1)\sinh r \cosh r$. The anomalous correlation $M_s$ arises because $b_{s,\mathrm{in}}$ mixes the original input operators $b_{\mathrm{in}}$ and $b_{\mathrm{in}}^\dagger$; the bath therefore acts as a squeezed thermal bath in the transformed frame. For the quadratures $q_s, p_s$, the noise diffusion entries are
\begin{equation}
D_{qq}=\gamma_b(2N_s+2M_s+1)=\gamma_b(2N_b+1)e^{2r},\qquad
D_{pp}=\gamma_b(2N_s-2M_s+1)=\gamma_b(2N_b+1)e^{-2r},
\end{equation}
with $D_{qp} = 0$ for a real parameter $r$. This yields the full block-diagonal diffusion matrix
\begin{equation}
D = \mathrm{diag}\Big[ \kappa_m (2N_m+1), \, \kappa_m (2N_m+1), \, \gamma_a (2N_a+1), \, \gamma_a (2N_a+1), \, \gamma_b (2n_{\text{th}}+1) e^{2r}, \, \gamma_b (2n_{\text{th}}+1) e^{-2r} \Big].
\end{equation}
In the low-temperature limit ($k_B T \ll \hbar \omega_m, \hbar \omega_a$) or under effective sideband cooling, the thermal occupations of the magnon and acoustic modes can be neglected ($N_m = N_a = 0$). In this regime, the noise from these modes is purely vacuum noise, reducing their diffusion blocks to $\kappa_m \mathbf{1}_2$ and $\gamma_a \mathbf{1}_2$.
\section{Stability analysis}
The stability of the steady-state solution is fundamentally tied to the roots of the characteristic polynomial associated with the drift matrix of the linearized system. By employing the Routh-Hurwitz stability criterion, we evaluate the real parts of the system's eigenvalues across the relevant parameter space. 
	\begin{figure}[tbhp]
		\centering
		\includegraphics[width=0.5\linewidth]{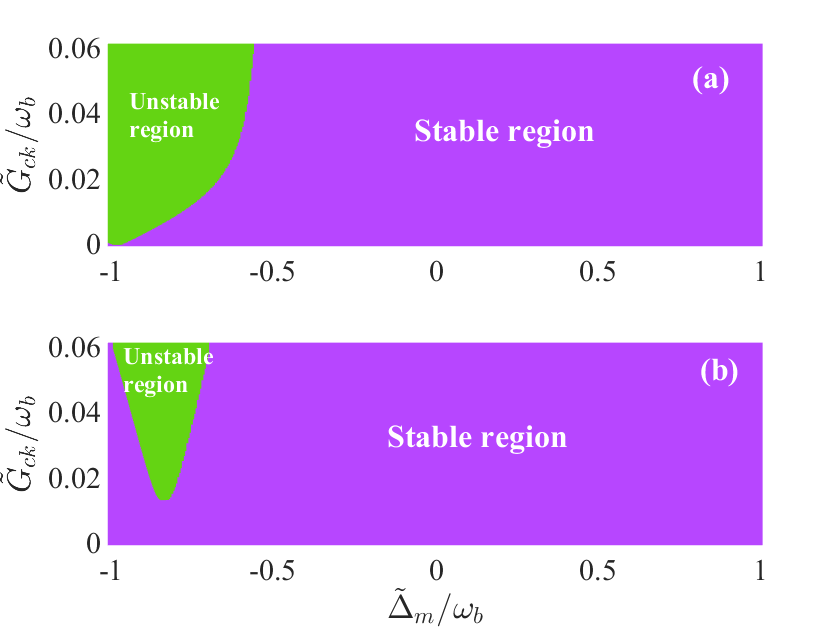}
		\caption{Stability boundaries in $(\tilde{G}_{kc}/\omega_b,\tilde{\Delta}_m/\omega_b)$ parameter space. The purple region is stable ($\lambda_{\max}<0$) and the green region is
			unstable ($\lambda_{\max}\geq0$), where the covariance matrix is not defined.
			(a) Without phonon hopping and MPA ($J_m/\omega_b=0$, $\chi/\omega_b=0$). (b) With phonon hopping and MPA ($J_m/\omega_b=0.1$, $\chi/\omega_b=0.5$).
			All other parameters are as in \cref{tab:parameters}.
		}
		\label{fig:stability_full}
	\end{figure}
	A stable operating regime requires that all eigenvalues satisfy $\lambda_{\max}<0$; otherwise, the system becomes dynamically unstable. In what follows, we investigate how key physical mechanisms namely, phonon hopping $(J_m)$ and mechanical parametric amplification ($\chi$) modify these stability boundaries within the $(\tilde{G}_{kc}/\omega_b,\tilde{\Delta}_m/\omega_b)$ parameter space. 

\cref{fig:stability_full} shows the stability boundaries in the $(\tilde\Delta_m,\tilde G_{ck})$
space. Instability occurs only for negative detuning,
$\tilde\Delta_m/\omega_b\lesssim-0.5$, i.e.\ for a blue-detuned drive, where the
two-mode-squeezing part of the coupling amplifies fluctuations. 
The unstable
region grows with $\tilde G_{ck}$ and is much smaller in the presence of phonon
hopping and parametric amplification strength (see \cref{fig:stability_full} (b)) than without them (see \cref{fig:stability_full} (a)). The operating point used in this work, $\tilde\Delta_m/\omega_b=0.5~ or~1$ and
$\tilde G_{ck}/\omega_b=0.05$, lies well within the stable regime. The stability boundaries in \cref{fig:stability_full} show strong agreement with the entanglement maps of \cref{fig:fig1,fig:fig2}. In \cref{fig:fig1,fig:fig2}, computed without phonon hopping, parametric amplification strength ($J_m/\omega_b=0$, $\chi/\omega_b=0$), and with  phonon hopping, parametric amplification strength ($J_m/\omega_b=0.1$, $\chi/\omega_b=0.5$), the white regions where no value is plotted coincide with the unstable region of \cref{fig:stability_full} (a,b). Consequently entanglement is evaluated only where $\lambda_{\max}<0$, demonstrating that the presence phonon hopping and parametric amplification expands the region of the $(\tilde\Delta_m,\tilde G_{ck})$ plane that supports entanglement.

\bibliography{Acoustic_Manuscript}
\end{document}